\documentclass[prb,twocolumn,superscriptaddress,%
preprintnumbers,nobibnotes]{revtex4}

\usepackage[dvips]{graphicx}%  JPEG PDF PNG EPS
\usepackage{textcomp}%  TS1 encoding \textdegree \textcelsius \textmu
\newcommand{\etal}{\textit{et al.}}
\newcommand{\ie}{i.e.}
\makeatletter
\def\Published@name{Published online }%
\def\ps@myheadings{%
    \let\@oddfoot\@empty\let\@evenfoot\@empty
    \def\@evenhead{\thepage\hfil\slshape\leftmark}%
    \def\@oddhead{{\itshape\rightmark}\hfil\thepage}%
    \let\@mkboth\@gobbletwo
    \let\sectionmark\@gobble
    \let\subsectionmark\@gobble
    }%
\makeatother
\def\rightmark{J.\ Appl.\ Phys. \textbf{101}, 033527 (2007)}
\begin{document}

\preprint{J.\ Appl.\ Phys. \textbf{101}, 033527 (2007)}

\title{Tip artifact in atomic force microscopy observations
of InAs quantum dots grown in Stranski--Krastanow mode}

\author{Ken-ichi~Shiramine}
\altaffiliation{Corresponding author}
\email{shira@eng.hokudai.ac.jp}
\author{Shunichi~Muto}
\affiliation{Department of Applied Physics,
Graduate School of Engineering,
Hokkaido University, Sapporo 060-8628, Japan}

\author{Tamaki~Shibayama}
\author{Norihito~Sakaguchi}
\author{Hideki~Ichinose}
\affiliation{Center for Advanced Research of Energy
Conversion Materials,
Hokkaido University, Sapporo 060-8628, Japan}

\author{Tamotsu~Kozaki}
\author{Seichi~Sato}
\affiliation{Division of Energy and Environmental System,
Graduate School of Engineering,
Hokkaido University, Sapporo 060-8628, Japan}

\author{Yoshiaki~Nakata}
\thanks{Present address: Nanoelectronics Collaborative Research Center,
University of Tokyo, 4-6-1 Komaba, Meguro, Tokyo 153-8505, Japan.}
\author{Naoki~Yokoyama}
\affiliation{Fujitsu Laboratories, Ltd., Atsugi 243-0197, Japan}

\author{Masafumi~Taniwaki}
\affiliation{Department of Environmental Systems Engineering,
Kochi University of Technology, Tosayamada, Kochi 782-8502, Japan}

\received{4 August 2006}
\accepted{4 December 2006}
\published{13 February 2007}

\begin{abstract}
The tip artifact in atomic force microscopy (AFM)
observations of InAs islands was evaluated
quantitatively. The islands were grown in the
Stranski--Krastanow mode of molecular beam epitaxy.
The width and height of the islands were
determined using transmission electron microscopy (TEM) and
AFM\@. The average $[\bar{1}10]$ in-plane width and
height determined using TEM excluding native oxide were
22 and 7\,nm, respectively; those determined using AFM
including the oxide were 35 and 8\,nm, respectively. The
difference in width was due to the oxide and the tip
artifact. The sizes including the oxide were deduced from
TEM observations to be a width of 27\,nm and a height of
6\,nm with correction for the thickness of the oxide. The
residual difference of 8\,nm between the width determined
using AFM and that determined using TEM including the oxide
was ascribed to the tip artifact. The results enable us to
determine the actual size of the islands from their AFM
images.
\ {\textcopyright} \textit{2007 American Institute of Physics} \
[DOI: 10.1063/1.2434806]
\end{abstract}

\maketitle

\section{Introduction}\label{intro}

Quantum dots have been expected to produce novel optical and
electronic devices because of their quantum size effect. Of
the devices, quantum-dot lasers will be realized
first.~\cite{Arakawa} Spintronic devices using quantum dots
will bring innovations in areas including quantum
information processing.~\cite{Ohno}

Islands are grown in the Stranski--Krastanow (S--K) mode of
molecular beam epitaxy (MBE)\@.~\cite{r38} The islands are
quantum dots: they have a nanometer size and can confine
carriers in them. The islands are suitable for devices since
they are coherent and show intense photoluminescence. They
are called self-assembled quantum dots.

The size of the islands is important information for
realizing the application of the islands to devices. It is
usually determined using transmission electron microscopy
(TEM) and atomic force microscopy (AFM)\@. TEM has the
advantage that it leads directly to the actual size of the
islands, although much time and effort are necessary to make
samples sufficiently thin for electron transmission. On the
other hand, AFM is a simple method that does not require
sample preparation, although the size determined from an AFM
image is not the actual size.

\begin{figure*}[tb]
\includegraphics[width=160truemm,clip]{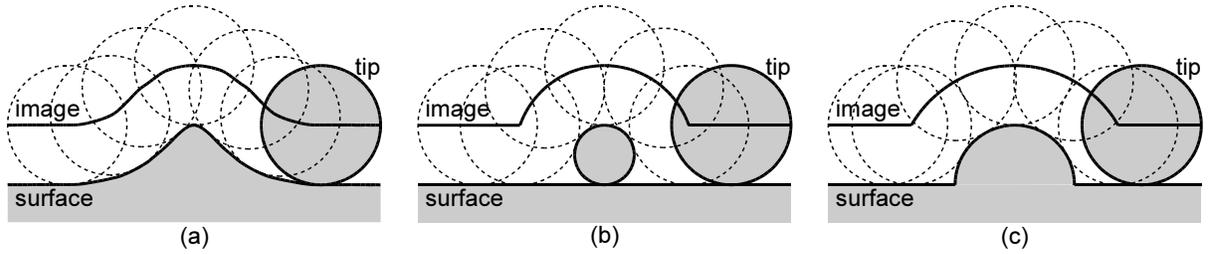}
\caption{Schematic views of AFM observations of (a) InAs
S--K island, (b) sphere, and (c) hemisphere on flat
surface. The AFM tip is depicted as a sphere having the same
radius as the curvature radius of the tip. The tip
artifact for the object having a steep surface, (b) and (c),
causes a larger broadening than that for the object having a
gentle surface, (a).\label{schema}}
\end{figure*}

In AFM observations, it is widely known that the \textit{tip
artifact} prevents the determination of actual
sizes~\cite{r1, r6, r8, r12, r14, r15, r18, r19, r23, r27,
r28, r30, r31, r40} (Fig.~\ref{schema}). Because of the
artifact, an AFM image does not represent the actual shape
of an object but a nonlinear convolution of the shape of the
object and the shape of the tip used. The in-plane width
of the object is hence overestimated, while the height is
determined correctly.~\cite{r6, r28, r37, r40} The artifact
occurs because (i) the tip has a finite size, and (ii) a point
of the tip other than its apex touches the
surface.~\cite{r1, r6, r14} The artifact is also called the
convolution, dilation, or tip effect.

The tip artifact in AFM observations can be suppressed by
(A) using a sharp tip or (B) the reconstruction of the
actual shape from an image. To suppress the artifact, some
research groups have attached a carbon-nanotube tip to a
commercially available probe,~\cite{r23, r31} (A)\@. Other
groups have tried to reconstruct the actual shape of an
object from its image by observing a sphere on a flat
surface as a reference,~\cite{r6, r12, r27, r28, r30, r31,
r40} (B)\@. A sphere is suitable as a reference for the
following reason. If we have or know two of the following
three things: an AFM image, the shape of the tip, and the
shape of the object, we can obtain the remaining
unknown.~\cite{r8} We can hence reconstruct the actual shape
of the object from its AFM image if we know the size of the
tip used. In addition, we can often determine the tip size by
observing a sphere on a flat surface, because the height of
the image is the diameter of the sphere and the width of
the image is the sum of the diameter of the sphere and the
diameter of the tip [Fig.~\ref{schema}(b)].

Geometric assumptions about an object produce an expression
for broadening due to the AFM tip artifact. Assume that
we observe a sphere of radius $r$ using a tip of radius
$R$ [Fig.~\ref{schema}(b)]. Simple consideration yields
$(R+r)^{2}=(w/2)^{2}+(R-r)^{2}$, where $w$ is a broadened
width of an image. It results in
\begin{equation}
 w=4\sqrt{Rr}.
 \label{eq1}
\end{equation}
In a similar way, if we assume that an object is a
hemisphere [Fig.~\ref{schema}(c)],
$(R+r)^{2}=(w/2)^{2}+R^{2}$ leads to
\begin{equation}
 w=2\sqrt{2Rr+r^{2}}.
 \label{eq2}
\end{equation}
A hemisphere is a more realistic model for the S--K islands
than a sphere because of the hemispherelike shape of the
islands.

In spite of much effort, only a few reports~\cite{r27,
r28, r31} have been published in which a tip artifact on the
nanometer scale was evaluated. The artifact in observations
of S--K islands has not yet been reported, although many
authors have pointed out its presence.~\cite{r32, r33, r34,
r35, r36, r37}

The determination of the broadening due to the tip artifact
gives us a method by which we can extract the actual size of
the S--K islands from their AFM images. The establishment of
this method provides a practical way to determine the actual
sizes of the islands of many samples. In fact, a research
group that makes samples of the islands using MBE usually
carries out hundreds of growth runs a year and observes all
the samples using AFM, whereas it cannot do the same using
TEM\@. This method using AFM, consequently, allows us to
determine actual sizes of the islands of hundreds of samples
a year.

In the application of the S--K islands to devices, the
actual size of islands embedded in a cap layer is
important. Note that the size of the embedded islands is
different from that of uncapped islands because of surface
segregation and interdiffusion. The relationship between the
sizes of the two kinds of island is beyond the scope of the
present study. Here, the relationship between the size of
the uncapped islands determined using AFM and that
determined using TEM is discussed as the first step.

In the present study, the size of the InAs S--K islands
grown by MBE was determined using TEM and AFM\@. The size
determined using TEM cannot be compared with that determined
using AFM because native oxide was present between the epoxy
and the crystalline (InAs and GaAs) in the TEM
sample. Observations using TEM yielded the size excluding
the oxide; observations using AFM yielded the size including
the oxide. We hence carefully corrected the size determined using TEM
for the thickness of the oxide to obtain a size
which we can compare with that determined using AFM\@.

\section{Experimental Method}\label{exp}

Indium arsenide S--K islands were grown by MBE on a GaAs
substrate at a low growth rate. A solid-source MBE with an
As$_{4}$ beam was used. The (001) $n^{+}$GaAs substrate was
fixed on a Mo block using In solder and was loaded into the
MBE system. The native oxide on the substrate surface was
desorbed by thermal etching. Then, a 250\,nm GaAs buffer
layer was grown at 610\,{\textcelsius} at a rate of
0.21\,{\textmu}m/h. Next, InAs of 2.4\,ML (monolayers) was
grown at 500\,{\textcelsius} at a rate of 0.006\,ML/s. The
temperature and As$_{4}$ pressure were maintained for
1.5\,min after the growth, and then the electrical heating
was turned off. No cap layer was grown. The uncalibrated
V/III ratios were 60 for InAs and 63 for GaAs. A TEM sample
and an AFM sample were cut from the same wafer.

We carried out (110) cross-sectional TEM observations.
First, two pieces of the wafer were fixed face to face using
epoxy. Then, a TEM sample was made from them by mechanical
dimpling and Ar ion thinning (Gatan, model 600) at 4\,kV and
15{\textdegree}. Field-emission TEM (JEOL, JEM-2010F) was
performed at 200\,kV\@. The width and height of the S--K
islands were determined from the TEM images. The width was
determined as the distance between the points where the
profile rose from the baseline.

\begin{figure}[tb]
\includegraphics[width=80truemm,clip]{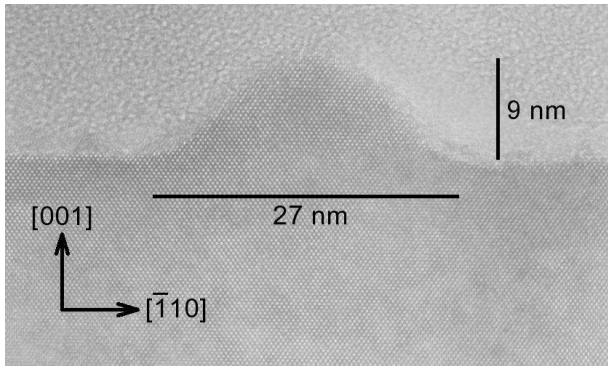}
\caption{Lattice image of InAs S--K island
obtained by (110) cross-sectional TEM observation. The width
and height of the island were determined to be 27 and
9\,nm.\label{TEM}}
\end{figure}

\begin{figure}[tb]
\includegraphics[width=50truemm,clip]{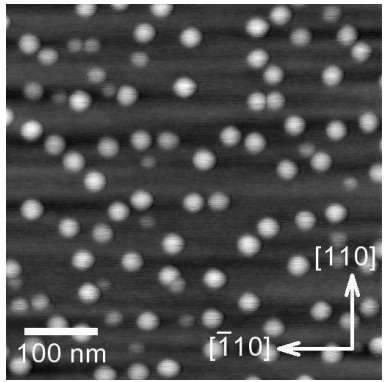}
\caption{AFM image of InAs S--K islands.\label{AFM}}
\end{figure}

Observations using AFM were carried out in air with a
MultiMode SPM/NanoScope IIIa (Veeco) operated
in the contact mode. A probe (Veeco, NP-20),
whose tip has a nominal curvature radius of 20\,nm, made of
Si$_{3}$N$_{4}$ was used. An image of $500\times
500\,\mathrm{nm}^{2}$ size was obtained by the $[\bar{1}10]$ scan
and was processed using \textsc{imagej} software.~\cite{ij} Details
of AFM observations and image processing are reported
elsewhere.~\cite{r39}

\section{Results and Discussion}

\begin{figure}[tb]
\includegraphics[width=80truemm,clip]{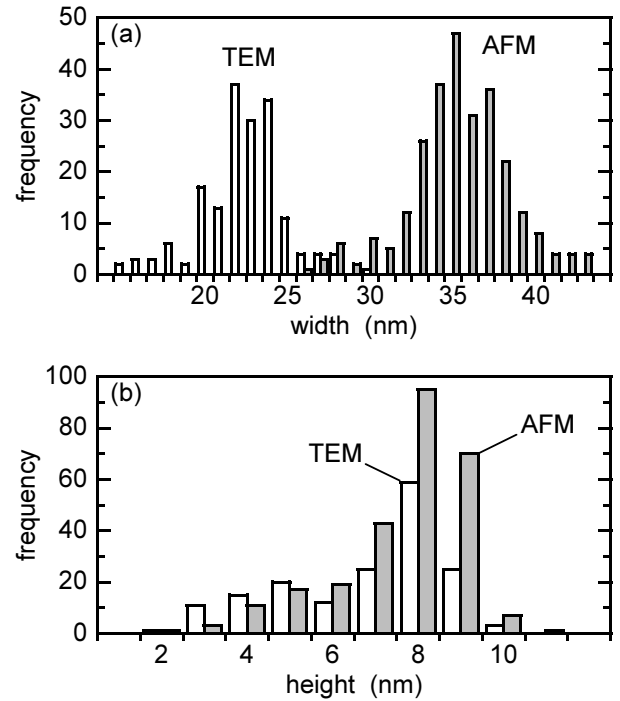}
\caption{Histograms of (a) $[\bar{1}10]$ in-plane width and (b)
height of InAs S--K islands obtained using TEM and AFM\@.
\label{histogram}}
\end{figure}

A cross-sectional TEM lattice image of an InAs S--K island
is shown in Fig.~\ref{TEM}. An AFM image of InAs S--K
islands is shown in Fig.~\ref{AFM}. The surface density of
the islands determined from the AFM images was $3.6\times
10^{10}\,\mathrm{cm}^{-2}$. The size of the islands was
determined from the TEM and AFM images. Shown in
Fig.~\ref{histogram} are the histograms of width and
height determined by TEM observations of 171 islands and AFM
observations of 267 islands. The average $[\bar{1}10]$
in-plane width and height determined from the TEM images
were 22 and 7\,nm, respectively; those determined from
the AFM images were 35 and 8\,nm, respectively.

The results suggest that the width determined using AFM was
13\,nm larger than that determined using TEM\@. This
difference was due to the native oxide and the tip artifact. As
will be shown, the size determined using TEM was corrected
for the thickness of the oxide. The width determined using
AFM was 8\,nm larger than that corrected after being
determined using TEM. The difference of 8\,nm was ascribed
to the tip artifact.

The size determined using TEM cannot be compared with that
determined using AFM owing to the native oxide. The oxide on the
surface was formed when the sample was removed from the MBE
equipment into air. The oxide of the TEM sample was embedded
in the epoxy (Sec.~\ref{exp})\@. Observations using AFM
yielded the island size including the oxide because AFM
images represent a topography. Observations using TEM, on
the other hand, yielded the island size excluding the oxide
because a diffraction contrast forms TEM images. Indeed,
since both were amorphous, the oxide could hardly be
distinguished from the epoxy in TEM observations. For the
above reasons, TEM observations could not yield the size of
the islands including the oxide embedded in the epoxy.

In spite of the above discussion, we succeeded in measuring
the size of the islands including the native oxide using TEM
in the course of the present study. In the TEM sample, there
was a region where the epoxy peeled off the surface. This
allowed us to determine the size including the oxide using
TEM\@. The sizes of 51 islands in the region were measured at
both (i) the surface of the oxide and (ii) the interface
between the oxide and the InAs. The thickness of the oxide
was 3\,nm on a flat surface. The average $[\bar{1}10]$
in-plane width and height determined in the former
method, (i), were 28 and 6\,nm, respecitvely; those
determined in the latter method, (ii), were 23 and
7\,nm, respectively (Table~\ref{tab1})\@. The latter sizes
agreed with the above sizes determined by TEM observations
of the 171 islands embedded in the epoxy. The width and
height determined in the former method, (i), were 5\,nm
larger and 1\,nm smaller than those determined in the latter
method, (ii), respectively.

\begin{table}
\catcode`?=\active \def?{\phantom{0}} \caption{[$\bar{1}$10]
in-plane width and height of the S--K islands determined
using TEM and AFM at the interface (I) and surface (S). The
number of the islands measured is shown in [\,], and the
standard deviation is shown in (\,).\label{tab1}}
\begin{ruledtabular}
\begin{tabular}{lccccc}
 & \multicolumn{2}{c}{TEM [51]} & \multicolumn{2}{c}{TEM
 [171]} & AFM [267] \\
 \cline{2-3} \cline{4-5} \cline{6-6}
 & I & S & I & S\footnotemark & S \\ \hline
 Width (nm) & 23(2) & 28(3) & 22(3) & 27 & 35(3) \\
 Height (nm) & ?7(2) & ?6(2) & ?7(2) & ?6 & ?8(1) \\
\end{tabular}
\end{ruledtabular}
\footnotetext{The fourth column was calculated from the third
 column. 5\,nm was added to the width, and 1\,nm was
 subtracted from the height.}
\end{table}

These differences between the sizes determined in both
methods are reasonable because the difference between the
shape of the surface and that of the interface is understood
as follows: The volume expands when the island surface is
oxidized. The oxide on the top of an island is thinner than
that on a flat surface because the surface area on the top
is enlarged by expansion; in contrast, the oxide on the tail
of an island is thicker than that on a flat surface because
the surface area on the tail is reduced.

We corrected the previously described TEM observations of
the width of 22\,nm and the height of 7\,nm for the
thickness of the native oxide using the above difference of
5\,nm in width and 1\,nm in height to obtain a width
of 27\,nm and a height of 6\,nm (Table~\ref{tab1})\@. The
corrected sizes should be obtained if the surface of many
islands were measured using TEM\@. They can hence be compared
with the sizes determined using AFM\@. The width
determined using AFM was, consequently, 8\,nm larger than
that corrected after being determined using TEM\@. The
difference in height was 2\,nm, and accordingly the heights
determined by both methods were considered to agree.

The residual difference in width of 8\,nm was ascribed to
the tip artifact in AFM observations. The broadening of
8\,nm is consistent with the nominal curvature radius,
{\ie}, 20\,nm, of the tip used. In fact, the artifact for
the islands having a gentle surface causes a smaller
broadening than the tip diameter (Fig.~\ref{schema}).

The broadening due to the tip artifact for the S--K islands
grown at a standard rate of $\sim$0.1\,ML/s is similar to
that for the islands grown at a low rate of 0.006\,ML/s
determined in the present study. For example, the islands
grown at the standard rate have a diameter of approximately
10\,nm (Fig.~2 of Ref.~\onlinecite{r39}), and their AFM
images show a diameter of approximately 20\,nm (Fig.~4 of
Ref.~\onlinecite{r39}). This difference is also ascribed to
the artifact.

The assumption that a S--K island has the shape of a hemisphere
permits us to estimate an image width, which is consistent
with that determined in the present study. In accordance
with Eq.~(\ref{eq2}), if we observe a hemisphere whose radius
$r$ is 8\,nm (Table~\ref{tab1}) using AFM with a tip whose
radius $R$ is 20\,nm, we should obtain an image having a
width $w$ of 39\,nm, which agrees with 35\,nm
(Table~\ref{tab1}) in the present results.

The broadening due to the tip artifact reported to date is
different from that determined in the present study. There
is, however, no inconsistency. Indeed, the difference in
broadening can be understood in terms of the shape of
objects, because the broadening depends on the
shape of objects.

We can compare the results of AFM observations of a
nanometer-sized sphere on a flat surface with those of the
present study. Xu and Arnsdorf observed gold colloidal
particles on mica using AFM and obtained an image with
11.7\,nm height and 44.4\,nm width.~\cite{r28} Using
Eq.~(\ref{eq1}), their results produce a tip radius $R$ of
21\,nm, which is close to that in the present study. Junno
{\etal}\ observed GaAs particles using AFM to evaluate the
artifact.~\cite{r40} At first, they observed a particle
whose image gave a width of 60\,nm. Then, they moved the
particle on a substrate by pushing it with the probe so that
it made contact with another particle to obtain, from an
image, a width of 90\,nm for the two particles touching each
other. Both actual diameter and the broadening were hence
determined to be 30\,nm. The above two results are
consistent with our results, whereas the broadening reported
is larger than ours, {\ie}, 8\,nm. In fact, an object with a
steeper surface induces a larger broadening owing to the
artifact [Figs.~\ref{schema}(b) and \ref{schema}(c)] than an
object with a gentle surface [Fig.~\ref{schema}(a)]. Having
a steeper surface, a sphere consequently induces a larger
broadening than the islands. Next, Chen {\etal}\ observed
gold nanoparticles using a tip made with a carbon nanotube
to obtain, from an AFM image, a width of 10\,nm.~\cite{r31}
The width of 10\,nm was the sum of a nanoparticle diameter
of 6\,nm and a tip diameter of 3.5\,nm; this is in
accordance with Fig.~\ref{schema}(b). The small broadening
is due to the small radius and a high aspect ratio of the
tip.

Merz {\etal}\ reported AFM observations of CdSe S--K
islands.~\cite{r37} They extrapolated the relation between a
height and a diameter of the islands to zero height to
determine a broadening of 23\,nm.

Using the broadening determined in the present study, we can
determine the actual size of the S--K islands by correcting
sizes obtained from AFM images. This is a practical way to
determine the actual sizes of the islands of many samples,
as discussed previously. The correction corresponds to a
reconstruction of an actual shape, (B) in
Sec.~\ref{intro}\@. By the way, an AFM image depends on the
size of the tip used: the employment of a probe of a
different size leads to a different image for an identical
object although the same type of probe is used. On the other
hand, one should not use AFM if the exchange of a probe
causes a complete change in an image. A probe having a large
tip is, actually, excluded as a ``bad probe'' in
observations.~\cite{r28} For this exclusion, observations of
an identical object result in almost the same image as long
as the same type of probe is used. The dependence of images
on tip size does not deteriorate the value of the correction
using the artifact. Some groups have determined a tip size
using a standard object;~\cite{r27, r41, r42} this procedure
allows us to confirm the tip size. Combination of the
procedure and the correction using the broadening assures
reliability of the correction.

In conclusion, the width of the InAs S--K islands
determined using AFM was overestimated by 8\,nm because of
the tip artifact. The results enable us to determine the
actual size of the islands from AFM observations. The
results of AFM observations on the islands reported to date
should be reevaluated considering the artifact.


\begin{thebibliography}{99}

\bibitem{Arakawa}
Y. Arakawa and H. Sakaki,
Appl. Phys. Lett. \textbf{40}, 939 (1982).

\bibitem{Ohno}
H. Ohno,
Science \textbf{281}, 951 (1998).

\bibitem{r38}
D. Leonard, M. Krishnamurthy, C.~M. Reaves, S.~P. Denbaars,
and P.~M. Petroff,
Appl. Phys. Lett. \textbf{63}, 3203(1993).

\bibitem{r28}
S. Xu and M.~F. Arnsdorf,
J. Microsc. \textbf{173}, 199 (1994).

\bibitem{r6}
P. Markiewicz and M.~C. Goh,
Langmuir \textbf{10}, 5 (1994).

\bibitem{r40}
T. Junno, K. Deppert, L. Montelius, and L. Samuelson,
Appl. Phys. Lett. \textbf{66}, 3627 (1995).

\bibitem{r14}
U. D. Schwarz, H. Haefke, P. Reimann, and
H.-J. G\"{u}ntherodt,
J. Microsc. \textbf{173}, 183 (1994).

\bibitem{r1}
M.~F. Tabet and F.~K. Urban~III,
Thin Solid Films \textbf{290-291}, 312 (1996).

\bibitem{r23}
K.~I. Hohmura, Y. Itokazu, S.~H. Yoshimura, G. Mizuguchi,
Y. Masamura, K. Takeyasu, Y. Shiomi, T. Tsurimoto,
H. Nishijima, S. Akita, and Y. Nakayama,
J. Electron. Microsc. \textbf{49}, 415 (2000).

\bibitem{r31}
L. Chen, C.-L. Cheung, P.~D. Ashby, and C.~M. Lieber,
Nano Lett. \textbf{4}, 1725 (2004).

\bibitem{r12}
J. Vesenska, R. Miller, and E. Henderson,
Rev. Sci. Instrum. \textbf{65}, 2249 (1994).

\bibitem{r27}
S. Xu and M.~F. Arnsdorf,
J. Microsc. \textbf{187}, 43 (1997).

\bibitem{r30}
H.~G. Abdelhady, S. Allen, S.~J. Ebbens, C, Madden, N,
Patel, C.~J. Roberts, and J. Zhang,
Nanotechnology \textbf{16}, 966 (2005).

\bibitem{r8}
D.~J. Keller and F.~S. Franke,
Surf. Sci. \textbf{294}, 409 (1993).

\bibitem{r18}
L. Hellemans, K. Waeyaert, F. Hennau, L. Stockman,
I. Heyvaert, and C. Van Haesendonck,
J. Vac. Sci. Technol. B \textbf{9}, 1309 (1991).

\bibitem{r19}
D. Keller,
Surf. Sci. \textbf{253}, 353 (1991).

\bibitem{r15}
P. Gr\"{u}tter, W. Zimmermann-Edling, and D. Brodbeck,
Appl. Phys. Lett. \textbf{60}, 2741 (1992).

\bibitem{r37}
J.~L. Merz, S. Lee, and J.~K. Furdyna,
J. Cryst. Growth \textbf{184/185}, 228 (1998).

\bibitem{r32}
S. Ruvimov, P. Werner, K. Scheerschmidt, U. G\"{o}sele,
J. Heydenreich, U. Richter, N.~N. Ledentsov,
M. Grundmann, and D. Bimberg,
Phys. Rev. B \textbf{51}, 14 766 (1995).

\bibitem{r35}
B. Junno, T. Junno, M.~S. Miller, and L. Samuelson,
Appl. Phys. Lett. \textbf{72}, 954 (1998).

\bibitem{r36}
I. Mukhametzhanov, Z. Wei, R. Heitz, and A. Madhukar,
Appl. Phys. Lett. \textbf{75}, 85 (1999).

\bibitem{r33}
P.~B. Joyce, T.~J. Krzyzewski, G.~R. Bell, T.~S. Jones,
S. Malik, D. Childs, and R. Murray,
Phys. Rev. B \textbf{62}, 10 891 (2000).

\bibitem{r34}
Q. Xie, J.~L. Brown, and J.~E. Van Nostrand,
Appl. Phys. Lett. \textbf{78}, 2491 (2001).

\bibitem{ij}
URL: \url{http://rsb.info.nih.gov/ij/}

\bibitem{r39}
K. Shiramine, S. Muto, T. Shibayama, H. Takahashi,
T. Kozaki, S. Sato, Y. Nakata, and N. Yokoyama,
J. Vac. Sci. Technol. B \textbf{21}, 2054 (2003).

\bibitem{r41}
S.~S. Wong, A.~T. Woolley, T.~W. Odom,
J.-L. Huang, P. Kim, D.~V. Vezenov, and C.~M. Lieber,
Appl. Phys. Lett. \textbf{73}, 3465 (1998).

\bibitem{r42}
J.~H. Hafner, C.-L. Cheung, T.~H. Oosterkamp, and
C.~M. Lieber,
J. Phys. Chem. B \textbf{105}, 743 (2001).

\end{thebibliography}
\end{document}